\documentclass[10pt,journal,compsoc]{IEEEtran}

\usepackage[nocompress]{cite}

\usepackage{url}
\usepackage{ragged2e}
\usepackage{epsfig}
\usepackage{graphicx}
\usepackage{amsmath,amssymb} 
\usepackage{mathptmx}      
\usepackage{subfigure}
\usepackage{algorithm}
\usepackage{algorithmicx}
\usepackage{algpseudocode}
\usepackage{graphics}
\usepackage{mathrsfs}
\usepackage{threeparttable}
\usepackage{color}
\usepackage[normalem]{ulem}
\usepackage{multirow}
\usepackage{float}
\usepackage{amsfonts}
\usepackage{bm}
\usepackage{array}
\usepackage[table]{xcolor}
\usepackage{colortbl}
\usepackage{pifont}
\usepackage{diagbox}
\usepackage{rotating}
\usepackage{booktabs}
\usepackage{overpic}
\usepackage{textcomp}
\usepackage{contour}
\usepackage{enumitem}
\usepackage{stfloats}
\usepackage{threeparttable}
\usepackage{makecell}
\usepackage[colorlinks=true,breaklinks=true,urlcolor=magenta]{hyperref}

\def\ie{\emph{i.e.}}

\def\etal{{\em et al.~}}

\definecolor{mygray1}{gray}{.75}

\def\ie{\textit{i.e.}}

\graphicspath{{./Imgs/}}
\DeclareGraphicsExtensions{.jpg,.pdf,.png}

\RequirePackage{silence}
\begin{document}

\title{SAM Meets Robotic Surgery: An Empirical Study in Robustness Perspective}

\author{
An Wang$^\dag$,~
Mobarakol Islam$^\dag$,~
Mengya Xu,~
Yang Zhang,~
Hongliang Ren$^*$\\
\IEEEcompsocitemizethanks{
\IEEEcompsocthanksitem An Wang, Yang Zhang, and Hongliang Ren are with the Department of Electronic Engineering, The Chinese University of Hong Kong, Hong Kong SAR, China.
\IEEEcompsocthanksitem Mobarakol Islam is with the Wellcome/EPSRC Centre for Interventional and Surgical Sciences (WEISS), Department of Medical Physics and Biomedical Engineering, University College London, UK.
\IEEEcompsocthanksitem Mengya Xu and Hongliang Ren are with the Department of Biomedical Engineering, National University of Singapore, Singapore.
\IEEEcompsocthanksitem Yang Zhang is with the School of Mechanical Engineering, Hubei University of Technology, China.
\IEEEcompsocthanksitem Corresponding author: Hongliang Ren (hlren@ee.cuhk.edu.hk).
\IEEEcompsocthanksitem An Wang and Mobarakol Islam are co-first author.
}
}

\IEEEtitleabstractindextext{%
\begin{abstract} \justifying
Segment Anything Model (SAM) is a foundation model for semantic segmentation and shows excellent generalization capability with the prompts. In this empirical study, we investigate the robustness and zero-shot generalizability of the SAM in the domain of robotic surgery in various settings of (i) prompted vs. unprompted; (ii) bounding box vs. points-based prompt; (iii) generalization under corruptions and perturbations with five severity levels; and (iv) state-of-the-art supervised model vs. SAM. We conduct all the observations with two well-known robotic instrument segmentation datasets of MICCAI EndoVis 2017 and 2018 challenges. Our extensive evaluation results reveal that although SAM shows remarkable zero-shot generalization ability with bounding box prompts, it struggles to segment the whole instrument with point-based prompts and unprompted settings. Furthermore, our qualitative figures demonstrate that the model either failed to predict the parts of the instrument mask (e.g., jaws, wrist) or predicted parts of the instrument as different classes in the scenario of overlapping instruments within the same bounding box or with the point-based prompt. In fact, it is unable to identify instruments in some complex surgical scenarios of blood, reflection, blur, and shade. Additionally, SAM is insufficiently robust to maintain high performance when subjected to various forms of data corruption. Therefore, we can argue that SAM is not ready for downstream surgical tasks without further domain-specific fine-tuning.  
\end{abstract}

}

\maketitle

\IEEEdisplaynontitleabstractindextext

\IEEEpeerreviewmaketitle

\IEEEraisesectionheading{
\section{Introduction}\label{sec:introduction}
}


\IEEEPARstart{S}{egmentation} of surgical instruments and tissue is a crucial challenge for robot-assisted surgery, and it is essential for instrument tracking and position estimation in surgical scenes. However, present deep learning models are frequently specialized to specific surgical sites, and their generalization capacity is restricted. As a result, establishing foundation models that can adapt to diverse surgical scenes and segmentation aims is critical in advancing robotic surgery~\cite{seenivasan2023surgicalgpt}. Recently, segmentation foundation models have made great progress in the field of natural image segmentation. The segment anything model (SAM)~\cite{kirillov2023segment}, which has been trained on more than 1 billion masks and has significant skills for creating accurate object masks based on prompts (e.g., bounding boxes, points, sentences), is the original and best-known segmentation foundation model. Several works have revealed that SAM can fail on common medical image segmentation tasks~\cite{deng2023segment,hu2023sam, he2023accuracy, ma2023segment}. This is reasonable and predictable given that SAM's training set consists primarily of natural image datasets with substantial edge information. Evaluating the performance of SAM on surgical scenes remains an underexplored field with scope to investigate further. In this report, we evaluate the generalizability of SAM under bounding box and point prompted settings on two public robotic surgery datasets. 

\section{Experiments}\label{sec:experiment}
In this study, two widely-used surgical instrument segmentation datasets, \ie, EndoVis17~\cite{allan20192017} and EndoVis18~\cite{allan20202018}, have been adopted in our experiments. Our evaluation involves three categories. Firstly, we have provided both quantitative and qualitative assessments on the promptable segmentation performance of SAM, with bounding boxes and single points, for binary and instrument-wise segmentation.
Besides, we comprehensively test the robustness of SAM with 18 types of data corruption at 5 severity levels by following~\cite{hendrycks2018benchmarking}. 
Moreover, we also examine SAM on its automatic mask generation in unprompted settings for surgical scene segmentation. 

\subsection{Experimental Settings}
\noindent\textbf{Datasets.}
We have employed two classical datasets in endoscopic surgical instrument segmentation, \ie, EndoVis17~\cite{allan20192017} and EndoVis18~\cite{allan20202018}. We follow the data splits in previous works~\cite{shvets2018automatic,gonzalez2020isinet,jin2019incorporating} to conduct the evaluation for SAM. 

\noindent\textbf{Prompts.} 
The original EndoVis datasets~\cite{allan20192017, allan20202018} do not have bounding boxes or point annotations. 
We have labeled the datasets with bounding boxes for each instrument, associated with corresponding class information. Additionally, regarding the single-point prompt, we obtain the center of each instrument mask by simply computing the moments of the mask contour. Since SAM~\cite{kirillov2023segment} only predicts binary segmentation masks, for instrument-wise segmentation, the output instrument labels are assigned inherited from the input prompts. 

\noindent\textbf{Metrics.} 
The IoU and Dice metrics from the EndoVis17~\cite{allan20192017} challenge\footnote{\url{https://github.com/ternaus/robot-surgery-segmentation}} is used. Specifically, only the classes presented in a frame are considered in the calculation for instrument segmentation.

\noindent\textbf{Comparison methods.}
We have involved several classical and recent methods, including the vanilla UNet~\cite{ronneberger2015u}, TernausNet~\cite{shvets2018automatic}, MF-TAPNet~\cite{jin2019incorporating}, Islam \etal\cite{islam2019real}, Wang \etal~\cite{wang2022rethinking}, ST-MTL~\cite{islam2021st}, S-MTL~\cite{seenivasan2022global}, AP-MTL~\cite{islam2020ap}, ISINet~\cite{gonzalez2020isinet}, TraSeTR~\cite{zhao2022trasetr}, and S3Net~\cite{baby2023forks} for surgical binary and instrument-wise segmentation. The ViT-H-based SAM~\cite{kirillov2023segment} is employed in all our investigations. Note that we cannot provide an absolutely fair comparison because existing methods do not need prompts during inference.

\begin{table*}[htbp]
  \centering
  \caption{\textbf{Quantitative comparison of different methods for binary and instrument segmentation on EndoVis17~\cite{allan20192017} and EndoVis18~\cite{allan20202018} datasets.} The adopted metric is the IoU (\%) in EndoVis17 Challenge~\cite{allan20192017}. The best and runner-up results are shown in bold and underlined, respectively. With bounding boxes as prompts, SAM~\cite{kirillov2023segment} shows superior performance compared with other methods. BBOX denotes for bounding box.}
    \begin{tabular}{cccccccc}
    \toprule
    \multirow{2}[2]{*}{Type} & \multirow{2}[2]{*}{Method} & \multirow{2}[2]{*}{Pub/Year(20-)} & \multirow{2}[2]{*}{Arch.} & \multicolumn{2}{c}{EndoVis17~\cite{allan20192017}} & \multicolumn{2}{c}{EndoVis18~\cite{allan20202018}} \\
\cmidrule{5-8}          &       &       &       & Binary IoU & Instrument IoU & Binary IoU & Instrument IoU \\
    \midrule
    \multirow{5}[2]{*}{Single-Task} & Vanilla UNet~\cite{ronneberger2015u} & MICCAI15 & UNet~\cite{ronneberger2015u}  & 75.44 & 15.80 & \uline{68.89} & - \\
          & TernausNet~\cite{shvets2018automatic} & ICMLA18 & UNet~\cite{ronneberger2015u}  & 83.60 & 35.27 & -     & 46.22 \\
          & MF-TAPNet~\cite{jin2019incorporating} & MICCAI19 & UNet~\cite{ronneberger2015u}  & 87.56 & 37.35 & -     & 67.87 \\
          & Islam \etal\cite{islam2019real} & RA-L19 & -     & 84.50 & -     & -     & - \\
          & ISINet~\cite{gonzalez2020isinet} & MICCAI21 & Res50~\cite{he2015deep} & -     & 55.62 & -     & 73.03 \\
          & Wang \etal~\cite{wang2022rethinking} & MICCAI22 & UNet~\cite{ronneberger2015u}  & -     & -     & 58.12 & - \\
    \midrule
    \multirow{5}[0]{*}{Multi-Task} & ST-MTL~\cite{islam2021st} & MedIA21 & -     & 83.49 & -     & -     & - \\
          & AP-MTL~\cite{islam2020ap} & ICRA20 & -     & \uline{88.75} & -     & -     & - \\
          & S-MTL~\cite{seenivasan2022global} & RA-L22 & -     & -     & -     & -     & 43.54 \\
          & TraSeTR~\cite{zhao2022trasetr} & ICRA22 & Res50~\cite{he2015deep} + Trfm~\cite{vaswani2017attention} & -     & 60.40 & -     & \uline{76.20} \\
          & S3Net~\cite{baby2023forks} & WACV23 & Res50~\cite{he2015deep} & -     & \uline{72.54} & -     & 75.81 \\
    \midrule
    \multirow{2}[0]{*}{Prompt-based} & SAM~\cite{kirillov2023segment} 1 Point & arxiv23 & ViT~\cite{dosovitskiy2021an}   & 53.88 & 55.96 & 57.12 & 54.30 \\
          & SAM~\cite{kirillov2023segment} BBox & arxiv23 & ViT~\cite{dosovitskiy2021an}   & \textbf{89.19} & \textbf{88.20} & \textbf{89.35} & \textbf{81.09} \\
    \bottomrule
    \end{tabular}%
  \label{tab:overall_res}%
\end{table*}%

\begin{table*}[htbp]
  \centering
  \caption{\textbf{Quantitative results on different splits of EndoVis17~\cite{allan20192017} and EndoVis18~\cite{allan20202018} datasets.}}
    \begin{tabular}{ccccccccc}
    \toprule
    \multirow{3}[4]{*}{Method} & \multicolumn{4}{c}{EndoVis17~\cite{allan20192017} (Seq9, Seq10)} & \multicolumn{4}{c}{EndoVis18~\cite{allan20202018} (Train + Val)} \\
\cmidrule{2-9}          & \multicolumn{2}{c}{Binary } & \multicolumn{2}{c}{Instrument} & \multicolumn{2}{c}{Binary } & \multicolumn{2}{c}{Instrument} \\
\cmidrule{2-9}          & IoU   & Dice  & IoU   & Dice  & IoU   & Dice  & IoU   & Dice \\
    \midrule
    SAM 1 Point & 58.99 & 71.04 & 57.19 & 67.68 & 59.81 & 70.89 & 57.43 & 65.91 \\
    SAM BBox & 90.41 & 94.86 & 86.58 & 91.32 & 87.15 & 92.64 & 81.08 & 86.53 \\
    \bottomrule
    \end{tabular}%
  \label{tab:additional_res}%
\end{table*}%

\begin{table*}[htbp]
  \centering
  \caption{\textbf{Quantitative results on synthetic EndoVis18~\cite{allan20202018} validation data with various corruptions.} 18 types of corruptions from 4 categories are employed to construct the synthetic datasets. Each corruption has 5 severity levels, and level 0 indicates clean data. The adopted metric is the IoU (\%) from EndoVis17 Challenge~\cite{allan20192017}. SAM~\cite{kirillov2023segment} encounters severe performance degradation in most cases but reacts differently to various types and levels of data corruption.
  }
    \resizebox{\textwidth}{!}{
    \begin{tabular}{cc|cccc|ccccc|ccccc|cccc}
    \toprule
    \multirow{2}[0]{*}{} & \multirow{2}[0]{*}{Severity} & \multicolumn{4}{c|}{Noise}    & \multicolumn{5}{c|}{Blur}             & \multicolumn{5}{c|}{Weather}          & \multicolumn{4}{c}{Digital} \\
          &       & Gaussian & Shot  & Impulse & Speckle & Defocus & Glass & Motion & Zoom  & Gaussian & Snow  & Frost & Fog   & Bright & Spatter & Contrast & Pixel & JPEG  & Saturate \\
    \midrule
    \multirow{6}[2]{*}{\begin{sideways}Binary\end{sideways}} & \multicolumn{1}{c}{0} & \multicolumn{18}{c}{89.35} \\
\cmidrule{2-20}          & 1     & 77.69 & 80.18 & 80.43 & 83.28 & 82.01 & 80.53 & 82.99 & 80.30 & 85.40 & 84.08 & 83.12 & 85.38 & 87.43 & 86.69 & 85.76 & 81.12 & 58.77 & 86.64 \\
          & 2     & 73.92 & 76.07 & 76.15 & 81.65 & 80.21 & 79.20 & 80.22 & 77.55 & 81.69 & 80.69 & 80.34 & 84.65 & 87.27 & 84.21 & 84.90 & 79.32 & 56.04 & 84.85 \\
          & 3     & 69.21 & 71.74 & 73.02 & 77.74 & 76.96 & 72.64 & 75.50 & 75.27 & 78.31 & 79.58 & 78.90 & 83.62 & 87.23 & 82.50 & 83.36 & 73.81 & 56.25 & 86.84 \\
          & 4     & 63.80 & 65.41 & 67.29 & 75.28 & 73.79 & 72.38 & 69.60 & 73.22 & 75.23 & 76.33 & 78.38 & 82.28 & 87.06 & 83.12 & 77.12 & 70.82 & 57.59 & 83.21 \\
          & 5     & 57.07 & 60.61 & 61.61 & 71.83 & 69.85 & 69.59 & 66.25 & 71.58 & 66.96 & 77.66 & 76.82 & 78.84 & 86.43 & 79.62 & 66.58 & 68.55 & 56.77 & 81.26 \\
    \midrule
    \multirow{6}[2]{*}{\begin{sideways}Instrument\end{sideways}} & \multicolumn{1}{c}{0} & \multicolumn{18}{c}{81.09} \\
\cmidrule{2-20}          & 1     & 69.51 & 71.83 & 72.25 & 74.82 & 73.64 & 72.13 & 74.33 & 71.41 & 76.79 & 75.40 & 74.42 & 76.82 & 79.16 & 78.24 & 77.17 & 72.94 & 54.86 & 78.27 \\
          & 2     & 66.06 & 68.09 & 68.53 & 73.19 & 71.74 & 71.02 & 71.46 & 68.85 & 73.15 & 72.13 & 71.65 & 76.14 & 79.00 & 75.54 & 76.22 & 71.55 & 52.23 & 76.61 \\
          & 3     & 62.01 & 64.44 & 65.89 & 69.75 & 68.74 & 64.97 & 67.13 & 67.12 & 70.08 & 70.97 & 70.21 & 75.01 & 78.90 & 73.70 & 74.67 & 66.83 & 51.63 & 78.39 \\
          & 4     & 57.28 & 59.12 & 61.03 & 67.82 & 65.87 & 64.87 & 62.15 & 65.18 & 67.23 & 68.43 & 69.79 & 73.73 & 78.73 & 74.24 & 69.48 & 63.99 & 51.88 & 74.91 \\
          & 5     & 51.56 & 55.16 & 55.86 & 64.76 & 62.43 & 62.23 & 59.26 & 63.96 & 60.60 & 69.33 & 68.32 & 70.45 & 78.19 & 70.72 & 61.14 & 61.79 & 51.01 & 73.35 \\
    \bottomrule
    \end{tabular}%
    }
  \label{tab:corruption}%
\end{table*}%

\begin{figure*}[!h]
  \centering
  \includegraphics[width=0.81\linewidth]{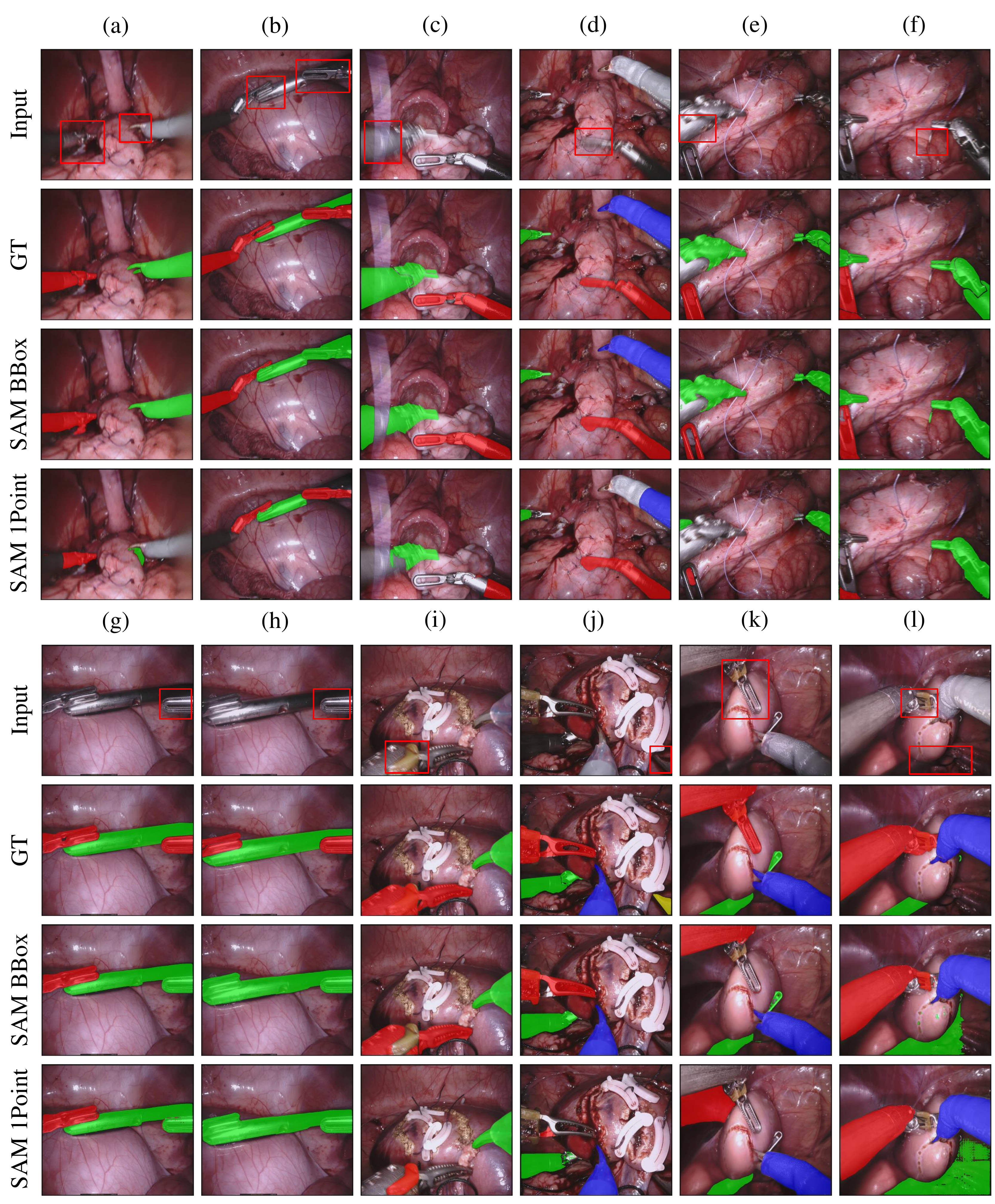}
  \caption{
    \textbf{Qualitative results of SAM~\cite{kirillov2023segment} on some challenging frames.} Red rectangles highlight the typical challenging regions which cause unsatisfactory predictions. Detailed analyses of each case are provided in the Result section. 
   }
  \label{fig:quali}
\end{figure*}

\begin{figure*}[!h]
  \centering
  \includegraphics[width=0.81\linewidth]{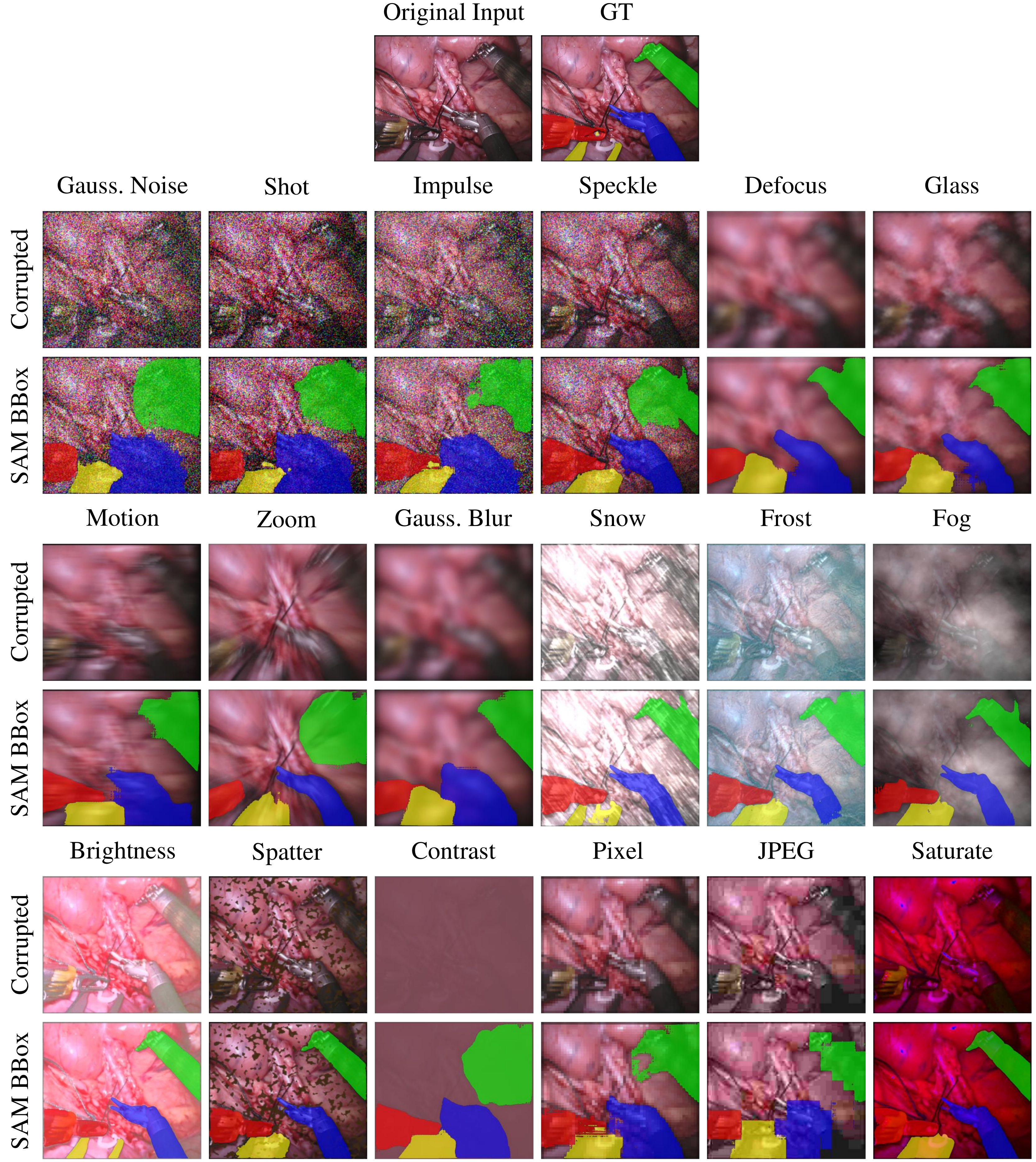}
  \caption{
    \textbf{Qualitative results of SAM~\cite{kirillov2023segment} under 18 data corruptions of level-5 severity.} Bounding boxes are adopted as the prompts. SAM~\cite{kirillov2023segment} can hardly maintain segmentation performance for most types of data corruption, showing its limited capacity for robustness. 
   }
  \label{fig:corrupt}
\end{figure*}

\subsection{Surgical Instruments Segmentation with Prompts}
\subsubsection{Implementation}
With bounding boxes and single points as prompts, we input the images to SAM~\cite{kirillov2023segment} to get the predicted binary masks for the target objects. Because SAM~\cite{kirillov2023segment} can not provide consistent categorical information. We compromise to use the class information from the bounding boxes directly. In this way, we derive instrument-wise segmentation while bypassing the possible errors from misclassifications, an essential factor affecting instrument-wise segmentation accuracy.  

\subsubsection{Results and Analysis}
SAM~\cite{kirillov2023segment}, as a foundational segmentation model, also exhibits great generalization capability in robotic surgery scenes.
As shown in Table~\ref{tab:overall_res}, with bounding boxes as prompts, SAM~\cite{kirillov2023segment} outperforms previous unprompted supervised methods in binary and instrument-wise segmentation on both datasets. Besides, when tested on different data splits settings, as presented in Table~\ref{tab:additional_res}, SAM~\cite{kirillov2023segment} still maintains its performance. These results demonstrate its superiority in generalizing to unseen endoscopic data under bounding box prompt of the surgical domains. With single points as prompts, SAM~\cite{kirillov2023segment} degrades a lot in performance, indicating its limited ability to segment surgical instruments from weak prompts. This reveals the performance of the SAM is closely rely on the heavy prompts.

For complicated surgical scenes, SAM~\cite{kirillov2023segment} still struggles to produce accurate segmentation results, as shown in columns (a) to (l) of Fig.~\ref{fig:quali}. Typical challenges, including shadows (a), motion blur (d), occlusion (b, g, h), light reflection (c), insufficient light (j, l), over brightness (e), ambiguous suturing thread (f), instrument wrist (i), and irregular instrument pose (k), all lead to unsatisfied segmentation performance.

\subsection{Robustness under Data Corruption}
\subsubsection{Implementation}
Referring to the robustness evaluation benchmark~\cite{hendrycks2018benchmarking}, we have evaluated SAM~\cite{kirillov2023segment} under 18 types of data corruptions at 5 severity levels with the official implementations\footnote{\url{https://github.com/hendrycks/robustness}}. Note that the \textit{Elastic Transformation} has been omitted to avoid inconsistency between the input image and associated masks. The adopted data corruption can be allocated into four distinct categories of \textit{Noise}, \textit{Blue}, \textit{Weather}, and \textit{Digital}. 

\subsubsection{Results and Analysis}
The severity of data corruption is directly proportional to the degree of performance degradation in SAM~\cite{kirillov2023segment}, as depicted in Table~\ref{tab:corruption}.
The resilience of the SAM algorithm~\cite{kirillov2023segment} may be influenced differently depending on the nature of the corruption present. However, in the majority of scenarios, the algorithm's performance diminishes significantly. Among the various types of corruption, \textit{JPEG Compression} and \textit{Gaussian Noise} have the greatest impact on segmentation performance, whereas \textit{Brightness} has a negligible effect. 
Figure~\ref{fig:corrupt} presents one exemplar frame in its original state, alongside various corrupted versions at a severity level of 5). We can observe that SAM~\cite{kirillov2023segment} still suffers great performance degradation in most cases.

\subsection{Automatic Surgical Scene Segmentation}
\subsubsection{Implementation}
Without prompts, SAM~\cite{kirillov2023segment} also supports automatic mask generation in  unprompted setting. For naive investigation of the automatic surgical scene segmentation results, we use the default parameters from the official implementation\footnote{\url{https://github.com/facebookresearch/segment-anything}} without further tuning. The colors of each segmented mask are randomly assigned because SAM~\cite{kirillov2023segment} only generates binary masks for each object. 

\subsubsection{Results and Analysis}
\begin{figure}[!h]
  \centering
  \includegraphics[width=\linewidth]{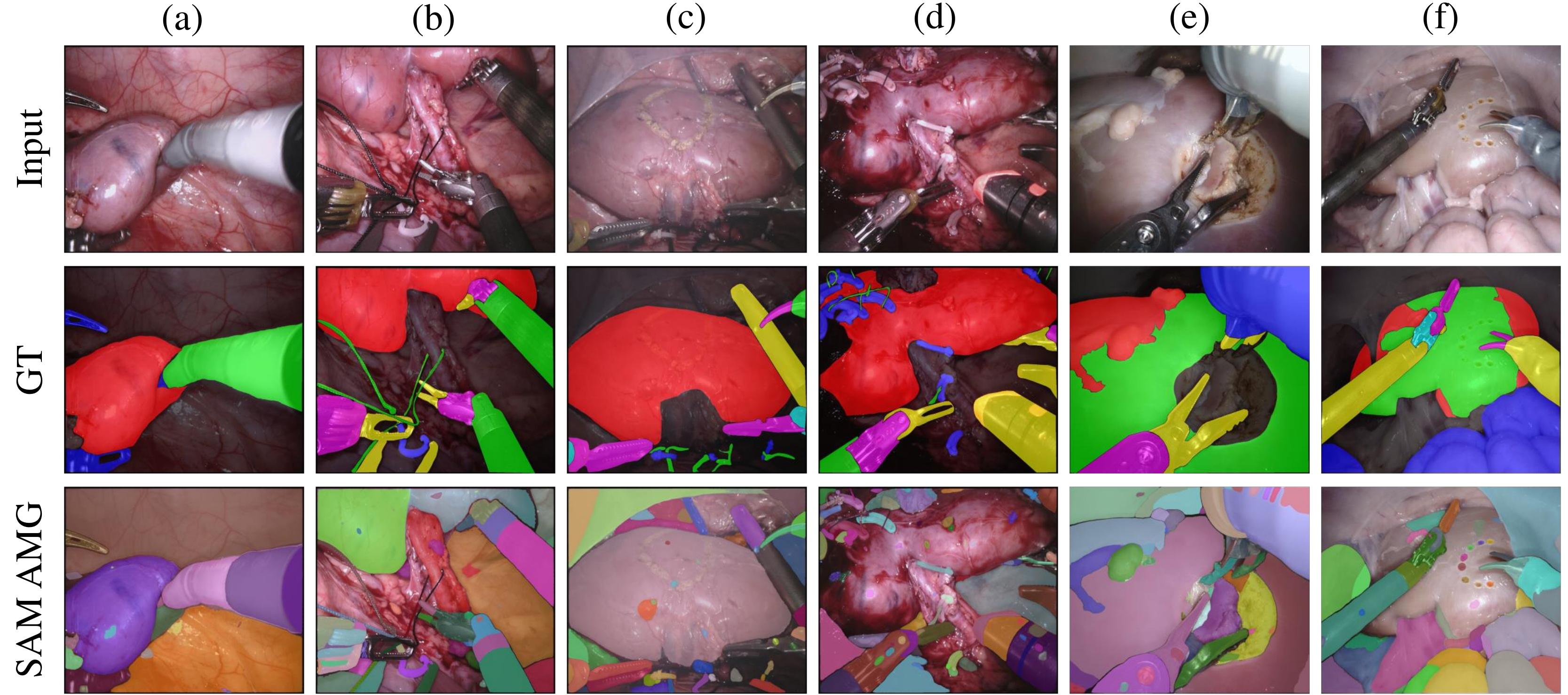}
  \caption{
    \textbf{Automatic mask generation (AMG) in  unprompted setting from SAM~\cite{kirillov2023segment} for EndoVis18~\cite{allan20202018} scene segmentation.} SAM~\cite{kirillov2023segment} cannot provide consistent categorical labels for each mask, and its comprehension of the proper surgical scene semantics is limited, leading to disconnected instrument articulating parts and fragmented anatomical tissue structures. 
   }
  \label{fig:amg}
\end{figure}

As shown in Fig.~\ref{fig:amg}, in surgical scene segmentation, SAM~\cite{kirillov2023segment} can produce promising results on simple scenes like columns (a) and (f). But it encounters difficulties when applied to more complicated scenes, as it struggles to differentiate between the entirety of instrument articulating parts accurately and to identify discrete tissue structures as interconnected units. As a foundation model, SAM~\cite{kirillov2023segment} still lacks comprehensive awareness of objects' semantics, especially in downstream domains like surgical scenes. 

\section{Conclusion}\label{sec:conclusion}


In this study, we explore the robustness and zero-shot generalizability of the SAM~\cite{kirillov2023segment} in the field of robotic surgery on two robotic instrument segmentation datasets of MICCAI EndoVis 2017 and 2018 challenges, respectively. 
Extensive empirical results suggest that SAM~\cite{kirillov2023segment} is deficient in segmenting the entire instrument with point-based prompts and unprompted settings, as clearly shown in Fig.~\ref{fig:quali} and Fig.~\ref{fig:amg}. This implies that SAM~\cite{kirillov2023segment}, although yielding surprising zero-shot generalization ability, can still not capture the surgical scenes precisely. 
Besides, it exhibits challenges in accurately predicting certain parts of the instrument mask when there are overlapping instruments or only with a point-based prompt. 
It also fails to identify instruments in complex surgical scenarios, such as blood, reflection, blur, and shade. 
Last but least, we extensively evaluate the robustness of SAM~\cite{kirillov2023segment} with a wide range of data corruptions. As indicated by Table~\ref{tab:corruption} and Fig.~\ref{fig:corrupt}, SAM~\cite{kirillov2023segment} encounters significant performance degradation in many scenarios. 

As a foundational segmentation model, SAM~\cite{kirillov2023segment} shows remarkable generalization capability in robotic surgical segmentation, yet it still suffers performance degradation due to downstream domain shift, data corruptions, perturbations, and complex scenes. To further improve its performance, a broad spectrum of evaluations and extensions, including fine-tuning, should be explored and developed.


\ifCLASSOPTIONcaptionsoff
  \newpage
\fi

{
\bibliographystyle{IEEEtran}
\bibliography{bibliography}
}


\vfill


\end{document}